\documentclass[usenatbib,usegraphicx]{mn2e}
\title[HD~44179's bipolar jet]{Geometry and velocity structure of HD~44179's bipolar jet}

\author[Joshua D. Thomas et al.]{Joshua~D.~Thomas,$^{1}$\thanks{E-mail:joshua.thomas@utoledo.edu} Adolf~N.~Witt,$^{1}$ 
Jason~P.~Aufdenberg,$^{2}$ J.~E.~Bjorkman,$^{1}$ \newauthor Julie~A.~Dahlstrom,$^{3}$  
L.~M.~Hobbs,$^{4}$  and Donald~G.~York$^{5}$\\
$^{1}$Ritter Astrophysical Research Center, The University of Toledo, Toledo, OH 43606\\
$^{2}$Physical Sciences Department, Embry-Riddle Aeronautical University, Daytona Beach, FL 32114\\
$^{3}$Department of Physics and Astronomy, Carthage College, Kenosha, WI 53140\\
$^{4}$Yerkes Observatory, The University of Chicago, Williams Bay, WI 53191\\
$^{5}$Department of Astronomy \& Astrophysics and The Enrico Fermi Institute, University of Chicago, Chicago, IL 60637}

\begin{document}

\date{Accepted for publication in MNRAS on 2012 December 20.  Received 2012 December 19; in original form 2012 November 9}

\pagerange{\pageref{firstpage}--\pageref{lastpage}} \pubyear{2012}

\maketitle

\label{firstpage}

\begin{abstract}

In this paper we analyse a set of 33 optical spectra, which were acquired with the ARCES echelle spectrograph (R = 38,000)
on the 3.5-m telescope at the Apache Point Observatory. We examine the H$_{\alpha}$ profile in each of these observations in 
order to determine the geometry and velocity structure of the previously discovered bipolar jet, which originates from the 
secondary star of HD~44179 located at the centre of the Red Rectangle nebula. Using a 3D geometric model we are able to determine 
the orbital coverage during which the jet occults the primary star. During the occultation, part of the H$_{\alpha}$ line profile 
appears in absorption.  The velocity structure of the jet was determined by modelling the absorption line profile using the Sobolev 
approximation for each orbital phase during which we have observations.  The results indicate the presence of a wide angle jet, 
likely responsible for observed biconical structure of the outer nebula.  Furthermore, we were able to determine a likely velocity 
structure and rule out several others.  We find that the jet is comprised of low-density, high-velocity, central region and a 
higher-density, lower-velocity, conical shell.

\end{abstract}


\section{Introduction}\label{paper2:sec:introduction}

The Red Rectangle (RR) is an X-shaped, biconical, protoplanetary nebula associated with the star HD~44179.  
Its moniker is derived from its rectangular appearance in red photographic plates \citep{cohen1975}, while its appearance in 
the blue plates is approximately circular.  The appearance of the RR in the red is dominated by its high intensity of extended 
red emission (ERE) \citep{schmidt1980,witt1990}, powered by the central source (HD 44179), which is a single line spectroscopic 
binary \citep{vanwinckel1995}.  The binary consists of a post-AGB primary star (0.8~M$_{\sun}$) with an effective temperature of 
roughly 8000~K, and a near solar mass main-sequence companion that has an accretion disk and jet (see Witt et al. 2009, and 
references therein).

The X-shaped structure corresponds  to the walls of a biconical low-density cavity within the spherical AGB outflow. The walls of 
the cavity appear bright where the line of sight is tangential to the cone.  The walls of the cavity are directly illuminated by 
the central source, which gives rise to the strong ERE \citep{schmidt1991, cohen2004, vijh2006}. The far UV photons 
(energies $> 10.5$~eV) necessary to excite the ERE \citep{witt2006} are produced in the hot accretion disk (T~$\sim$~17,000~K) of 
the secondary star \citep{witt2009}.

The RR's seemingly unique appearance can be attributed to its orientation on the sky with respect to our view from Earth. 
The central binary is obscured by a near-edge-on optically-thick circumbinary disk with an inclination angle of 86{\degr} as 
measured by \citet{bujarrabal2005} or 85{\degr} as measured by \citet{thomas2011}.  Therefore, the central source can only be 
observed via scattering over the top and bottom of the circumbinary disk. This indirect line of sight causes us to view the binary 
at an effective inclination angle of 35{\degr} \citep{waelkens1996}. 
The effective inclination is an average angle, as the orbital motion produces a variation of $\pm$~0.8{\degr}. 
Since the inferred properties of the system are dependent upon 
knowing the correct angle for our line of sight to the primary star, all such derived properties are somewhat uncertain.

Significant research has been conducted on the morphology of the extended nebula. 
The most likely agent for producing the morphology is a high speed bipolar jet launched
from the accretion disk around the secondary star. The bipolar jet proceeds to carve 
a low density cavity in the more slowly expanding spherical cloud of material ejected 
during the primary star's AGB stage. This scenario for shaping such 
bipolar nebulae was proposed by \citet{morris1981,morris1987}. Such a high speed jet has been observed in the RR via H$\alpha$ 
absorption \citep{witt2009}. Studies of the H$\alpha$ spectra reveal that the velocity of the 
jet material ranges between a minimum of 150~km~s$^{-1}$ \citep{koning2011} and a maximum of 560~km~s$^{-1}$ \citep{witt2009}. 

Recent work by \citet{velazquez2011} and \citet{koning2011} have used
the binary model to explain other features of the morphology of the RR. 
\citet{velazquez2011} posit a narrow precessing jet that originates from the secondary star 
with the goal of producing the ladder-like rungs that appear in the outer nebula.
Conversely, \citet{koning2011} employed a wide angle jet to evacuate the biconical cavity.  Using a scattered light model 
(with the biconical cavity devoid of dust) they show that the ladder rungs could be ring-like structures that are observed in other 
bipolar nebulae, which in the case of the RR, when projected onto the plane of the sky, appear as ladder rungs. \citet{soker2005} 
demonstrates that special physics is not required to shape the RR, and suggests intermittent jets give rise to the ladder like 
structures in the RR.

The goal of this paper is explore the geometry and kinematics of the jet using an analysis of the H$\alpha$ spectra as a function 
of orbital phase. Is the jet narrow or wide? Specifically, we demonstrate that the opening angle of the jet is roughly equal to the 
geometric opening angle of the X-shaped structure of the RR.  Therefore, our data are more consistent with the model of 
\citet{koning2011}.

In this paper, we first introduce the framework in which the data will be discussed.  The orbit and the physical model for the jet
 are described in {\S}~\ref{paper2:sec:model}.  We then present the observations in {\S}~\ref{paper2:sec:observations}, and the 
measurements in {\S}~\ref{paper2:sec:measurement}.  Calculations based on a numerical model and a discussion are presented in 
{\S}~\ref{paper2:sec:modeljet}. We present our conclusions in {\S}~\ref{paper2:sec:conclusions}.

\section{Model}\label{paper2:sec:model}

A 3D geometric model of the RR system is necessary for interpreting our measurements (to be discussed in 
{\S}~\ref{paper2:sec:measurement}).  The principal components of this system are: the primary star, the secondary star and 
its accretion disk, the bipolar jet, the circumbinary disk, and the spherical outflow. We approximate the jet as a bi-cone centred
 on the secondary star, and include the orbital solution of \citet{thomas2011} for calculating the positions of the various objects 
as a function of orbital phase.

\subsection{Geometry}\label{paper2:sec:model:geometry}
                                                    
The inclination angle, $i$, of the RR is 85{\degr}. The distance to the RR remains uncertain, but we adopt the
value of 710~pc derived by \citet{menshchikov2002} for our calculations. The outer radius of the
circumbinary disc is 1850~au \citep{bujarrabal2005}, while the radius of the central cavity in the circumbinary disk is estimated to 
be 14~au \citep{menshchikov2002}. The thickness of the circumbinary disk is determined to be approximately 100~au 
\citep{menshchikov2002,roddier1995, osterbart1997,bujarrabal2005}.  The opening angle (angle from cavity wall to cavity wall) 
of the biconical cavity is 40{\degr} just above the circumbinary disk, while at larger distances it expands from 62{\degr} to 
80{\degr} \citep{cohen2004}.  For modeling purposes we use 70{\degr} (corresponding to a 35{\degr} half opening angle), as was 
used in the work of \citet{waelkens1996}.

\begin{figure}
\begin{center}
\includegraphics[width=.95\linewidth]{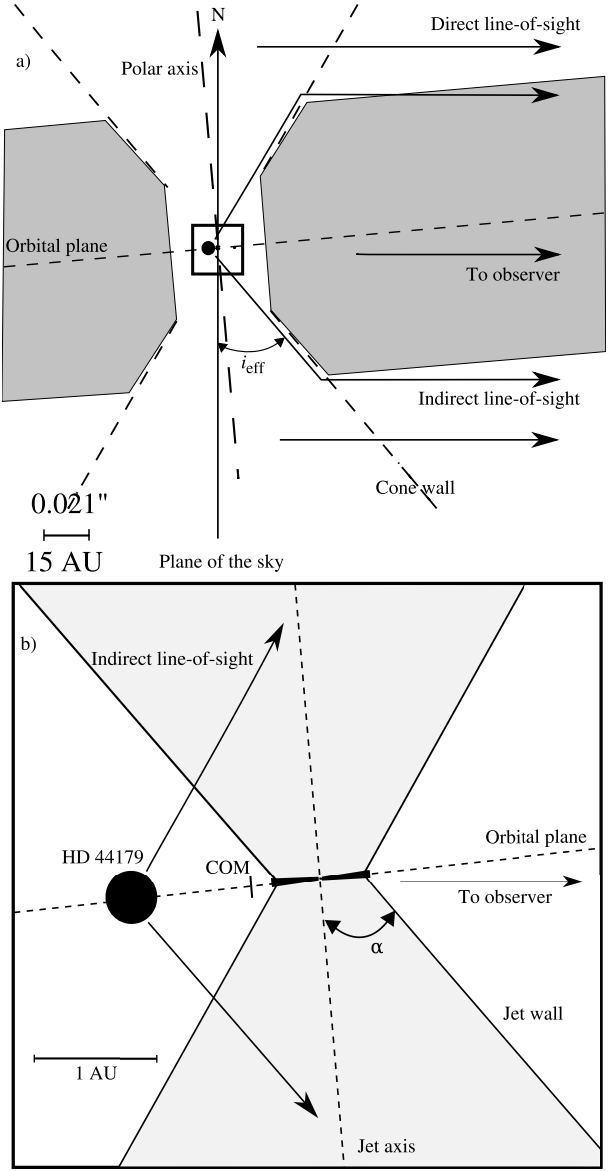}
\caption[Geometrical model.]{Geometrical model of the RR.  Panel (a) illustrates the overall geometry of the RR as seen from the plane of the sky. The height of the circumbinary disk and the inner radius are to-scale.  The binary at the centre appears 10 times larger than the scale of panel (a).  The outer edge of the circumbinary disk (1850~au) is not shown.  The cone walls indicated are meant to correspond to the X-shaped structure seen in the RR.  The polar axis shown is perpendicular to the orbital plane, which is tilted with respect to the direction of the observer by 5{\degr}  ($i =$ 85{\degr}). The outer nebula is seen via the direct line of sight, while the binary is seen under at the effective inclination, $i_{\mathrm{eff}}$ = 35{\degr}, via the indirect line of sight. Panel (b) is an enlargement of central portion of panel (a).  Panel (b) shows the indirect line of sight passing through the jet at the phase of superior conjunction (SC), $\phi$~=~0.59 (see Figure \ref{fig2}). The 
size of the primary star and the separation
between the primary and secondary are shown to-scale with 1~au as indicated. The accretion disk is shown around the secondary, with the jet emanating from above and below the hot accretion disk. The fiducial mark between the stars corresponds to the centre of mass (COM). }
\label{fig1}
\end{center}
\end{figure}

Figure \ref{fig1} illustrates the innermost region of our model for the RR. The outer radius of the accretion disk ($\sim$0.3~au) 
is set to be the distance of the L1 Lagrange point from the secondary star when the binary components are at periastron. The size of 
the primary star (radius $\sim$ 0.21~au) and the orbital separation of the binary components are shown to-scale. Notice that the 
direct line of sight (direction to the observer) is blocked by the optically thick circumbinary disk. Consequently, the primary star 
is viewed indirectly (at an effective inclination angle, $i_{\mathrm{eff}} =$ 35{\degr}) via scattering over the top and bottom 
of the circumbinary disk.  As a result, the indirect line of sight passes through the material in the jet above and below the 
accretion disk when the secondary passes in front of the primary at superior conjunction.  We will refer to the size of the 
jet by the half opening angle, $\alpha$, or the angle between the wall of the conical jet and the jet's axis, as shown in Figure 
\ref{fig1}.

As discussed in \citet{thomas2011}, emission features can originate anywhere in the RR, whereas absorption
features must arise from material in the line of sight to the primary star. Therefore, when studying absorption features one 
must use $i_{\mathrm{eff}} =$ 35{\degr}.  Furthermore, the absorbing material 
most likely lies between the star and the inner edge of the circumbinary disk. Our focus for this study will be on the periodically variable blue-shifted H$\alpha$ absorption feature attributed to the jet \citep{witt2009}.

\subsection{Orbit}\label{paper2:sec:model:orbit}
  
\begin{figure}
\includegraphics[width=\linewidth]{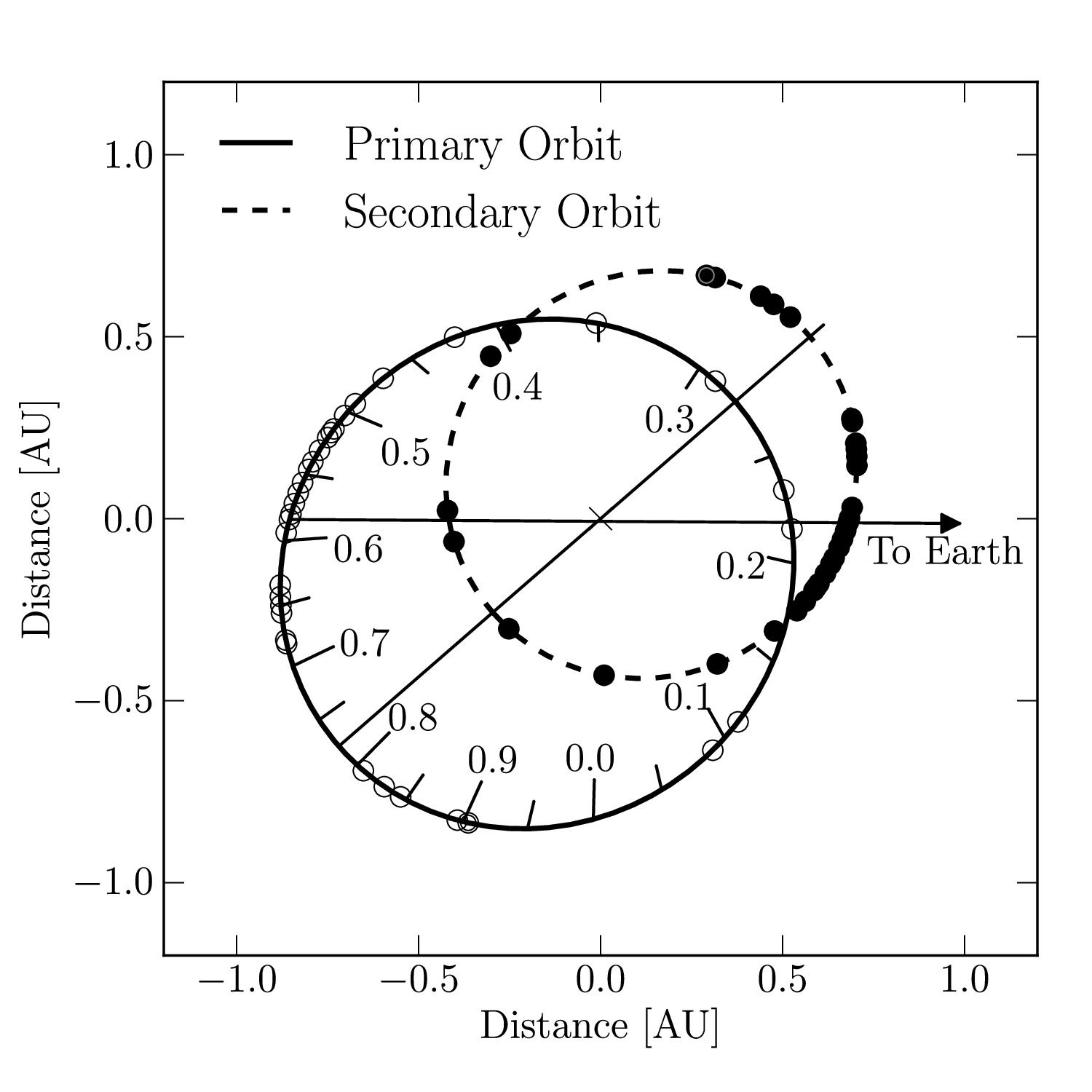}
\caption[Orbital plane diagram.]{Orbital plane diagram. The orbit
of the primary is shown by the solid ellipse, while the dashed ellipse
shows the orbit of the secondary. The orbital phases are indicated
on the orbit of the primary. The primary star moves in the direction
of increasing phase numbers, anticlockwise. Phase zero was chosen
to be consistent with \citet{waelkens1996}; it corresponds to JD 2448300.
The locations of the
stars for each of our observations are shown by the circles on the orbits.  The secondary is always opposite the centre of mass (COM), which is located at (0,0).  The locations of apastron and periastron are at the intersection of the orbit with the antipodal line. Inferior conjunction (IC), phase $\phi$~=~0.21, occurs where the direction vector to Earth (5{\degr} below the plane of the page) intersects the primary's orbit on the Earth side of the COM.  The phase of superior conjunction (SC), phase $\phi$~=~0.59, occurs where the Earth vector intersects the primary's orbit on the far side of the COM.}
\label{fig2}
\end{figure}

The orbit of the binary has been studied previously by \citet{waelkens1996} and \citet{thomas2011}, the orbital period was found to 
be $\sim$~317~days. The system is receding from Earth at roughly 19~km~s$^{-1}$.
In all of our data the observations are phase folded
using the orbital period of 317~days with a zero phase point $t_{0} =$~JD 2448300, which was chosen to correspond roughly to minimum in the radial velocity curve \citep{waelkens1996}. In this paper all orbital phases
indicate the location of the primary. The location of the secondary
is always opposite relative to the centre of mass. To aid in the interpretation of the data and model calculations, Figure \ref{fig2} shows the orbits of both stars.
We will frequently refer to the following key orbital locations: inferior conjunction (IC) when the primary is between the observer and the secondary, which occurs at
phase $\phi$~=~0.21; superior conjunction (SC) when the secondary is between the observer and the primary ($\phi$~=~0.59);
apastron ($\phi$~=~0.79); and periastron ($\phi$~=~0.29).

\section{Observations}\label{paper2:sec:observations}

The high velocity biconical jet is observed via absorption in the H$\alpha$ line profiles. Since the original study of the jet by \citet{witt2009} 16 new spectra have been obtained in order to study the geometry of the jet, with emphasis on the orbital phases during which the material in the jet lies in the line of sight to the primary star.  The data set now contains a total of 33 observations, the locations of which are shown in Figure \ref{fig2}. 

\subsection{Details of the Observations}\label{paper2:sec:observations:details}

The spectra were acquired at the Apache Point Observatory (APO) with
the ARCES echelle spectrograph \citep{wang2003} and the 3.5-m telescope.
The spectrograph has a resolving power of R~=~38,000, producing a velocity
resolution of 8~km~s$^{-1}$. The signal-to-noise ratio of the spectra in the H$\alpha$ region is
roughly 1000.  Each observation is the average of two spectra acquired consecutively on the same night. The dates of observation were chosen for the purpose of studying the periodic variability of the RR. The
data reduction has remained unchanged since \citet{thorburn2003} and \citet{hobbs2004}, and 
the complete data set is described fully in \citet{thomas2011}.

\begin{figure*}
\includegraphics[width=\linewidth]{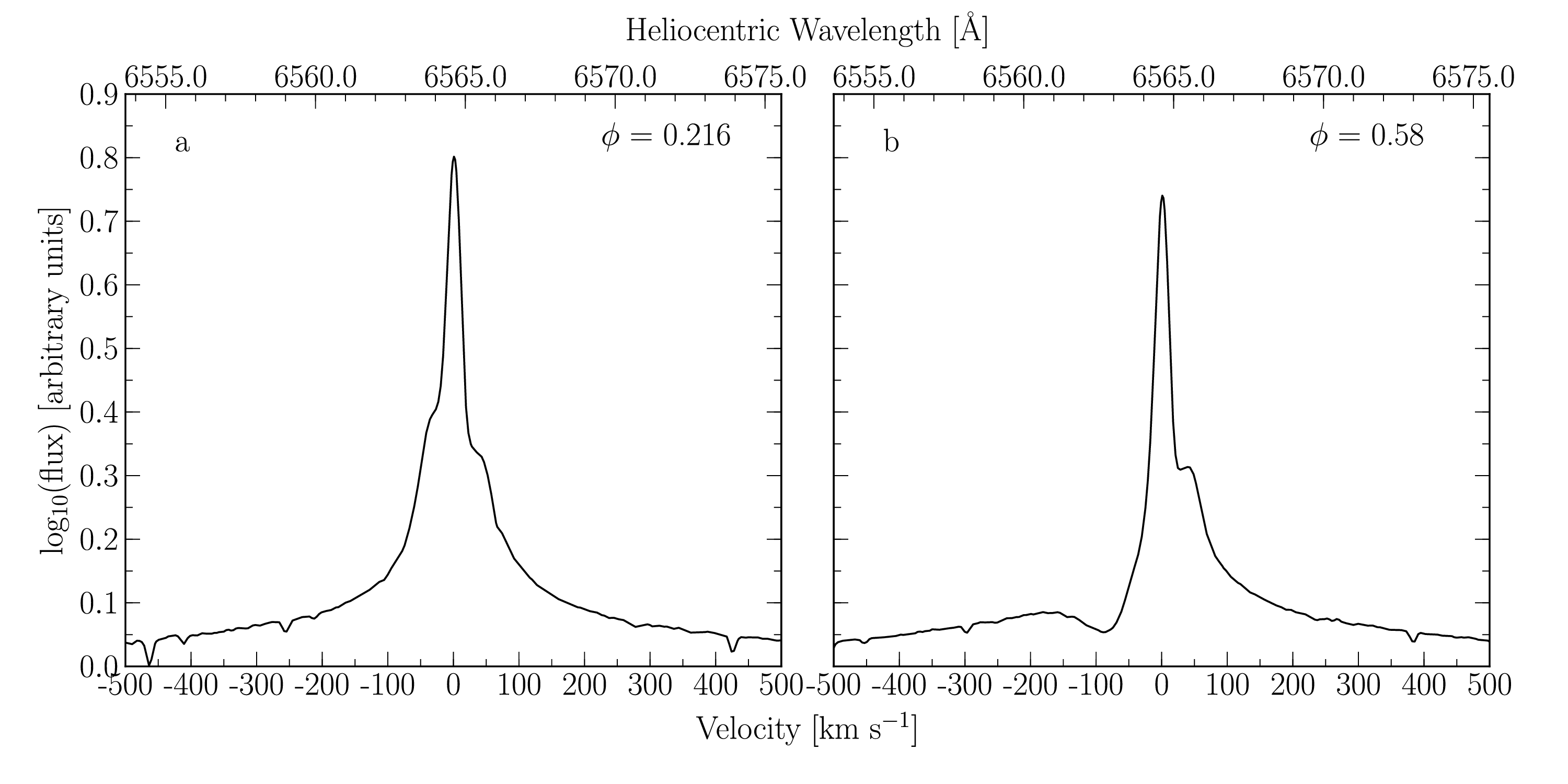}
\caption[Representative H$\alpha$ spectra.]{Representative H$\alpha$ spectra.  The log$_{10}$ of the flux is plotted on the vertical axis, as a result the continuum level is at log$_{10}(1)$ = 0.  This scaling was done to deemphasize the height of the narrow emission peak. The continuum-normalised spectra have been corrected for the underlying
photospheric absorption and its intrinsic broadening. Panel (a) shows
the most symmetric spectrum, which occurs at IC. Panel (b) shows a spectrum
at phase $\phi$~=~0.58, close to SC, with its strong blue-shifted absorption.
For the relative position of the stars at these phases see Figure
\ref{fig2}. The lower horizontal axis indicates the centre
of mass velocity in km~s$^{-1}$, to convert to heliocentric velocity add 19~km~s$^{-1}$.
}
\label{fig3}
\end{figure*}

The H$\alpha$ line profiles, shown in Figure \ref{fig3}, have several distinct features. First, there is a narrow, symmetric, stationary emission peak that has a FWHM $\sim$20~km~s$^{-1}$
\citep{jura1997, hobbs2004, witt2009}. Second, there is a (lower amplitude) broad emission plateau (FWHM of nearly 200~km~s$^{-1}$).  Third, there is an asymmetric blue-shifted absorption feature on the blue side of the emission plateau that appears shortly after the phase of IC and becomes strongest around the time of SC, see panel b of Figure \ref{fig3}.  There are also broad emission wings that extend to $\pm$~600~km~s$^{-1}$. The narrow emission feature has been ascribed to material in the central
H\textsc{ii} region \citep{jura1997, witt2009, thomas2011}. The likely source of ionising photons is the UV produced in the inner regions of the secondary's accretion disk \citep{witt2009}. The width of the narrow emission feature
is similar to the orbital velocities for the innermost circumbinary
material, which suggests that this material is in circumbinary orbit. The broad emission plateau and extended emission wings are likely due to either the material in the jet, and/or material in the accretion disk.  Since emission can originate from anywhere in the system it is harder to pinpoint its origin. The absorption feature has been attributed to absorption by material in the jet \citep{witt2009}. It is this absorption feature we shall focus on.

\subsection{Photospheric subtraction}\label{paper2:sec:observations:photo_subtraction}

Analysing the H$\alpha$ profiles in detail to study the jet requires that the observed spectral profiles
have the contribution of the primary's photosphere removed. The NLTE stellar atmosphere code, \textsc{phoenix}, version 
15.04.00E \citep{hau1992, hau1993, hau1995, allardhau1995, baron1996, hau1996, hau1997, baronhau1998, allard2001, hau2001}, 
was used to calculate the expected H$\alpha$ absorption line profile of the primary star. Other model lines produced by 
\textsc{phoenix} are much narrower than the observed lines \citep{thomas2011}; this feature likely applies to the H$\alpha$ line 
as well. Therefore, the subtraction of the \textsc{phoenix} model line from the observed line profiles was carried out using the subtraction
technique discussed in \citet{thomas2011}. This technique removes
the underlying photospheric absorption line by accounting for the
intrinsic broadening that seems to be present in all unblended atomic
absorption lines in the spectra of the RR \citep{hobbs2004, thomas2011}. The work of \citet{grinin2006} and \citet{grinin2012} attributes similar broadening to light scattering off moving grains. The model line profile is broadened by a broadening kernel derived from the observed characteristic photospheric absorption profiles. Examples of the
subtracted spectra are shown in Figure \ref{fig3};
these same spectra are presented in their unsubtracted form in Figure 5 of
\citet{witt2009}.

\section{Measurements}\label{paper2:sec:measurement}

\begin{figure}
\includegraphics[width=\linewidth]{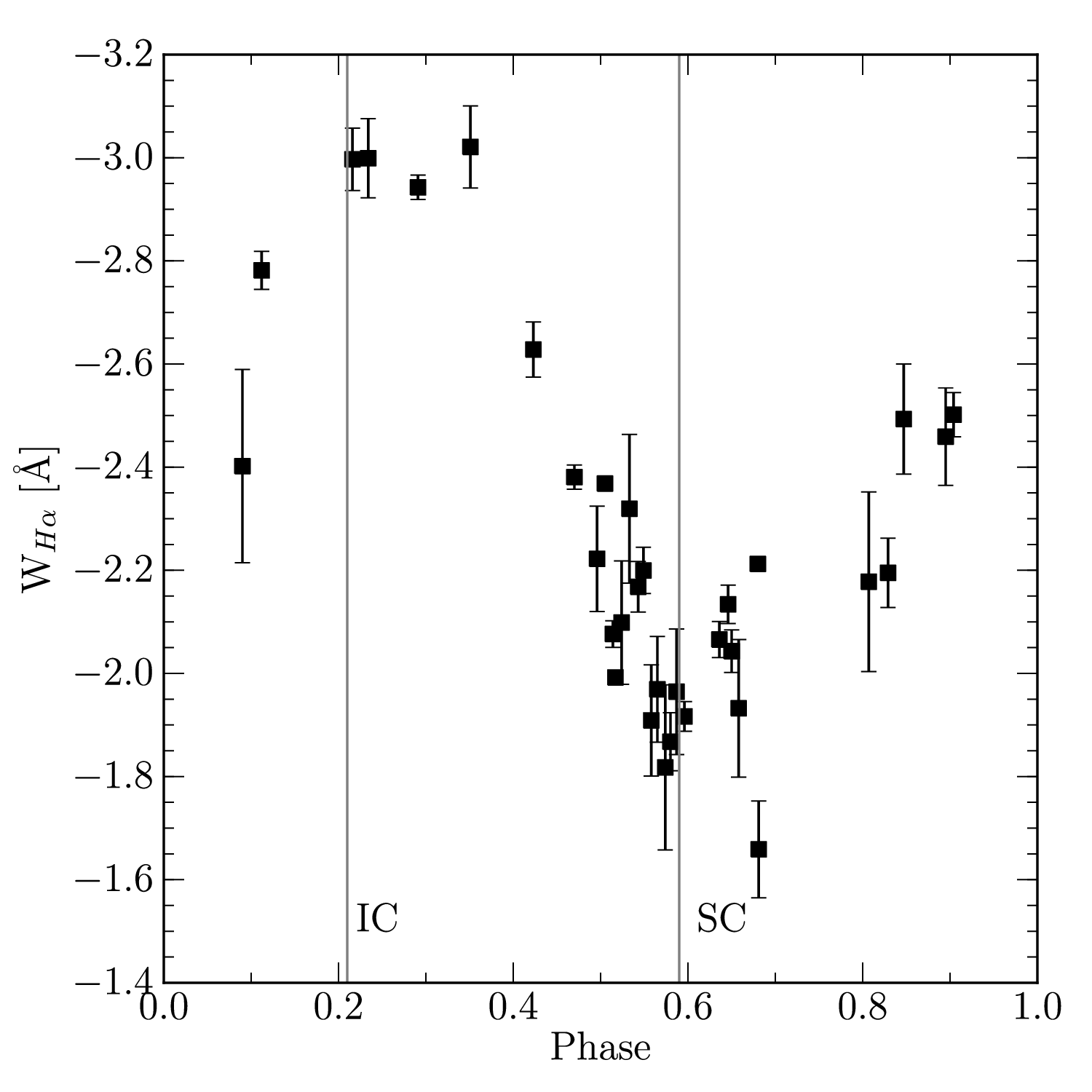}
\caption[Equivalent width measurements.]{Equivalent width measurements of the blue wing of the H$\alpha$ profile.  Larger negative values indicate stronger emission, and correspondingly less absorption by the jet.  The
error bars are one standard deviation for the average of two measurement
techniques. The vertical lines correspond to the phases of IC and SC, see Figure \ref{fig2}. The absorption maximum of the jet occurs near SC, while the minimum of the absorption occurs near IC.}
\label{fig4}
\end{figure}

In order to study the jet geometry, the equivalent width of the blue wing of the broad emission plateau of the H$\alpha$ profile, 
$W_{\mathrm{\lambda}}$(H$\alpha$)~=~$W_{\mathrm{H\alpha}}$, was measured as a function of orbital phase. The orbital phase dependence of these measurements helps to constrain $\alpha$, the half opening angle of the jet.  The emission equivalent width was measured for each photosphere-subtracted spectrum, from $-$27~km~s$^{-1}$
to $-$600~km~s$^{-1}$, so as to not include the effects of
the central narrow H$\alpha$ emission feature.  The results are shown in Figure \ref{fig4}.  Emission equivalent widths are typically defined to be negative, in Figure \ref{fig4} we see that the maximum H$\alpha$ equivalent width measured is roughly $-3$~{\AA}.  As a function of phase we see that the amount of emission decreases just after IC, reaching a minimum near SC after which it increases again as the primary star again approaches IC. The observed decrease corresponds to more absorbing material in the indirect line of sight, thus lowering the amount of emission observed. Since the equivalent width measurements remain negative, the amount of absorption never exceeds the amount of emission. The scatter in the data is due to uncertainties in the continuum level and residual telluric features. The equivalent widths were measured by hand and with a script, each data point is the average of these two values and the error bars represent one standard deviation.

\begin{figure}
\includegraphics[width=\linewidth]{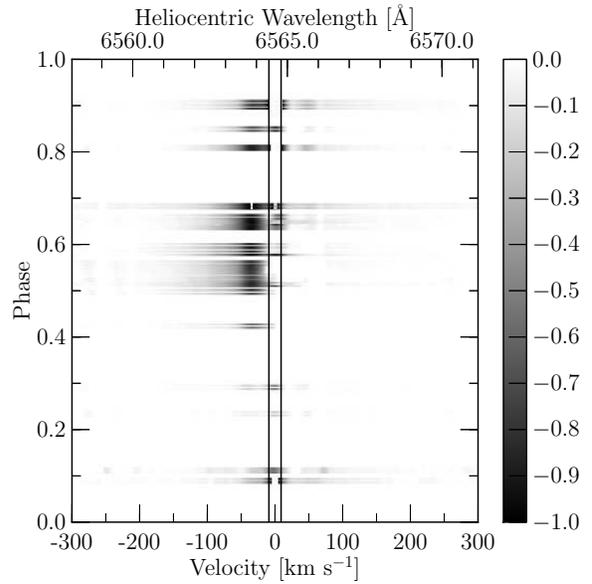}
%dynamical_data_color
\caption[H$\alpha$ dynamical spectrum]{Dynamical spectrum of the H$\alpha$ absorption feature. Each spectrum shown is the difference (observed~$-$~template).  The continuum corresponds to 0.00 on the linear grey-scale bar, absorption corresponds to negative numbers. The vertical lines correspond to the FWHM of the narrow emission peak, any subtraction here is inaccurate due to variations in the intensity of the narrow emission peak.}
\label{fig5}
\end{figure}

Another way to visualise the characteristic absorption feature of the jet is to construct a difference spectrum by subtracting the spectrum that shows no sign of the absorbing material.  The photosphere-subtracted spectrum which occurs at IC, phase $\phi$~=~0.21, was taken to be the template spectrum due to its symmetric observed profile (see panel a of Figure \ref{fig3}), and because this is the phase during which the emission equivalent width is least affected by the presence of the jet (i.e., the equivalent width is most negative), see Figure~\ref{fig4}. The template spectrum was subtracted from every photosphere-subtracted spectrum; the results are presented as a dynamical spectrum for all 33 observations in Figure \ref{fig5}. The systematic variation in the strength of the absorption as a function of orbital phase naturally agrees with the equivalent width measurements.  The majority of the absorption occurs within roughly $-$200~km~s$^{-1}$, while the 
underlying emission extends to approximately $-$600~km~s$^{-1}$.
The velocity at which the absorption maximum occurs, $V_{\mathrm{abs \ max}}$, is relatively constant, roughly $-$35~km~s$^{-1}$, while the velocity of the blue edge of the absorption feature, $V_{\mathrm{blue \ edge}}$, is variable and has a maximum value of roughly $-$200~km~s$^{-1}$.

\section{Modeling the Jet}\label{paper2:sec:modeljet}

In this section our goal is to determine the jet half opening angle, $\alpha$ (the angle between the conical jet axis and the wall of the conical jet, see Figure \ref{fig1}).  To do so we will employ two observational diagnostics that depend on the jet geometry.  The first is the absorption equivalent width, which measures the amount of absorbing material in the jet. The second is the dynamical absorption spectrum, which measures the velocities in the jet.

\subsection{Modeling the equivalent width measurements}\label{paper2:sec:modeljet:eqw}

From the observed phase coverage ($\phi$~=~0.4 to $\phi$~=~0.9), where the primary star is occulted to varying degrees by the jet
 (Figure \ref{fig4} and Figure \ref{fig5}), we can say that $\alpha$ must be large enough that the jet attenuates light from the 
primary star for most of the orbit. This implies that $\alpha$ is near but does not exceed the effective inclination of the line of 
sight to the primary star. If $\alpha$ were larger than $i_{\mathrm{eff}}$ there would always be some occultation of the primary 
star.

\begin{figure}
\includegraphics[width=\linewidth]{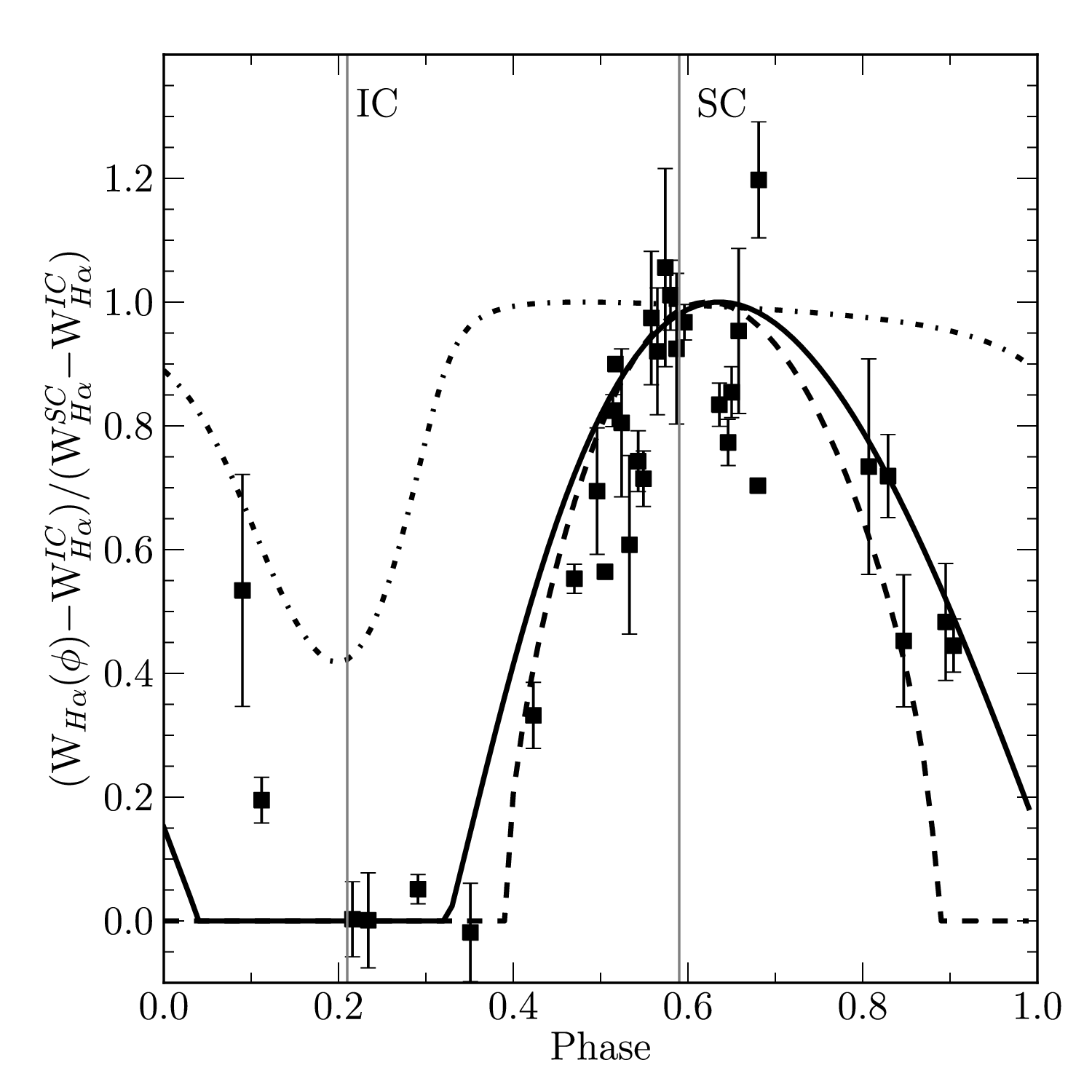}
\caption[Equivalent Width Difference and Model Curves]{The data points are the normalised difference in the equivalent width measurements for
the blue-shifted absorption feature of the H$\alpha$ profile, shown in arbitrary units.  The curves correspond to the normalised geometric path length through
the jet for different values of $\alpha$, also in arbitrary units. The dashed,
solid, and dash-dot curves correspond to $\alpha$ = 33{\degr},
35{\degr}, and 37{\degr} respectively. The horizontal axis corresponds to the orbital phase. The vertical lines correspond
to IC and SC, see Figure \ref{fig2}.
}
\label{fig6.eps}
\end{figure}

It is useful to look at the variation in the equivalent width of the absorption feature, instead of the emission feature.  We used 
the change in equivalent width with respect to the symmetric spectrum at IC.  The change in equivalent width produced by the 
absorption feature was calculated by subtracting the equivalent width at IC from each equivalent width measurement 
(W$_{H\alpha}(\phi) - $W$_{H\alpha}^{IC}$). The results were normalised for comparison to the normalised model curves 
(to be discussed below) and are shown in Figure \ref{fig6.eps}.  In the figure we see that as a function of phase the line 
of sight starts to pass through the jet just after IC. The line of sight reaches absorption maximum near SC, and trails off 
toward zero absorption as the primary star again approaches IC.

The path length through the jet was used to approximate the equivalent width, $W_{\mathrm{\lambda}}$, which on the linear portion 
of the curve of growth is proportional to the column density, $N$. At constant density, $N$ is proportional to the geometric 
path length, $l$, through the jet. This simplistic model allows us to calculate the orbital phase dependence of the normalised 
path length through the jet at various values of $\alpha$, for comparison to the equivalent width measurements.  Model 
calculations for three values of $\alpha$ (33{\degr}, 35{\degr}, and 37{\degr}) are over-plotted on the normalised equivalent 
width measurements in Figure \ref{fig6.eps}. The selected curves are calculated for an indirect line of sight angle of 
$i_{\mathrm{eff}}=35${\degr}, the ``effective'' inclination angle. The model curves fit surprisingly
well for such a simple model.   The results indicate that the best fitting model is such that
 $\alpha \approx $~35{\degr}~=~$i_{\mathrm{eff}}$.  Since the value of $i_{\mathrm{eff}}$ is uncertain, the path length curves were 
calculated for effective inclination angles of 30{\degr}, 40{\degr}, and 50{\degr}.  In all cases, the results were 
such that the value of $\alpha$ required to match the measurements was, respectively; 30{\degr}, 40{\degr}, and 50{\degr}. 
Therefore, we can state that $\alpha$ must approximately equal the effective inclination angle for the system, regardless 
of the assumed effective inclination. 

\begin{figure}
\includegraphics[width=\linewidth]{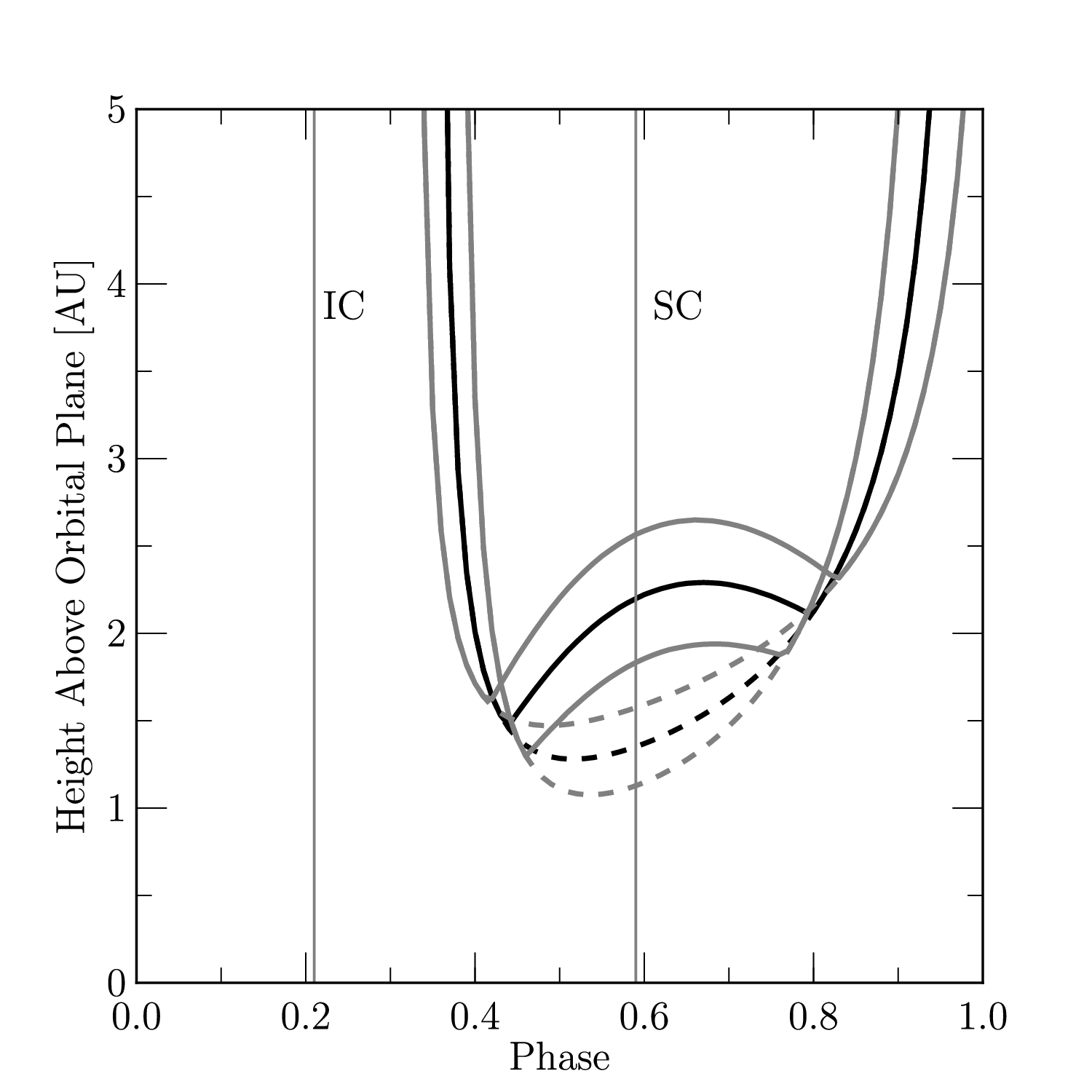}
\caption[Height above the orbital plane.]{Height above the orbital plane at which the indirect line of sight
passes through the jet as a function of phase.  Three lines of sight are shown.  The black curves are for the line of sight to the centre of the star.  The upper and lower grey curves correspond, respectively, to the upper and lower limbs of the star.  The long dashed curves correspond to where the lines of sight enter the jet;  the solid curves correspond to their intersection with the plane in the plane containing the axis of the jet. The vertical lines correspond to IC and SC, see Figure \ref{fig2}.
 }
\label{fig7}
\end{figure}

\citet{witt2009} noted that the equivalent width measurements are asymmetric about the phase of SC. This asymmetry was interpreted as a sign of material trailing the jet. However, we find that trailing material is not required to explain the observed asymmetry in the orbital phase dependence of the absorption feature. The curves are asymmetric due to the eccentricity of the orbit ($e \sim 0.3$) \citep{thomas2011}. The eccentricity of the orbit causes the indirect lines of sight to the primary star to pass through different heights above the orbital plane as a function of phase. 
Using our 3D model we calculate the height above the orbital plane for lines of sight to the upper and lower limbs of the star, as well as the centre of the primary star (see Figure \ref{fig7}).  The large size of the primary star  spreads out the curves for lines of sight to the leading and trailing limbs of the star as well; however, the change in phase coverage is minor (about $\pm 0.025$ in phase), the curves are otherwise similar. The resulting path lengths ($l =$~exit point (45~au) $-$ entry point) are also asymmetric about SC. Therefore, we need not invoke trailing material to explain the observations. Any such trailing material would be of significantly lower density than the jet, and would have a significantly smaller effect on the observed spectra. If present, this material may simply contribute to a more or less uniform background present in all spectra, and would be effectively removed in the subtraction of the template spectrum.

\subsection{Modeling Jet Velocities}\label{paper2:sec:modeljet:jetvel}

Here we use the dynamical absorption spectrum (Figure \ref{fig5}) to determine the jet kinematics by fitting the observed velocities. We will focus on the observed velocity of the blue edge of the absorption feature, $V_{\mathrm{blue \ edge}}$, and the observed velocity of the absorption maximum $V_{\mathrm{abs \ max}}$.  We will use the majuscule, $V$, to refer to velocities in the data, and the minuscule, $v$, to denote computed velocity values. For our calculations below, we used $\alpha$~=~$i_{\mathrm{eff}}$~=~35{\degr}.

We modelled a constant velocity jet, an accelerating jet, and a jet with a latitudinal velocity structure. The constant velocity model has a velocity of 200~km~s$^{-1}$.  The accelerating jet had only a radial dependence to the velocity of the form $v(r) = v_{0} \left(r/r_{norm}\right)$, where $v_{0}$ is the velocity along the jet axis (set to 200~km~s$^{-1}$), $r$ is the radial distance from the centre of mass of the secondary star, and $r_{norm}$ is the normalisation radius in this case we used 1~au.  In all the models we assume the outflow is purely in the radial direction, centred on the secondary star.

\begin{figure}
\includegraphics[width=\linewidth]{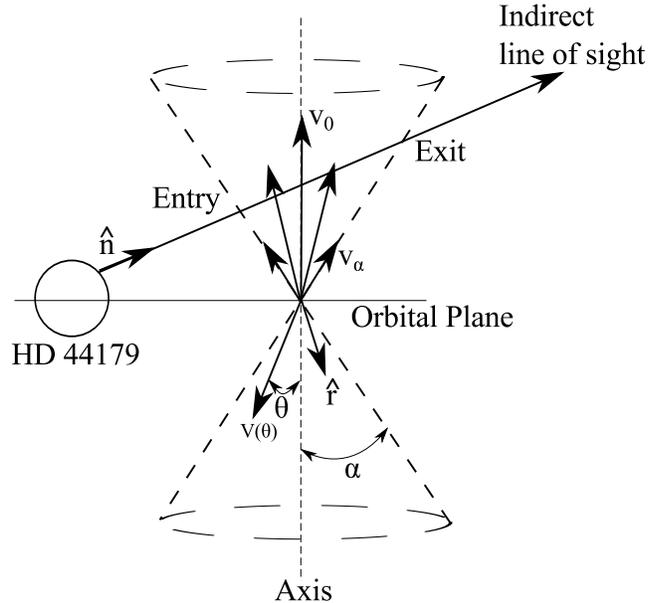}
\caption[Cone coordinates]{Simplified jet diagram illustrating the velocity structure described in the text.  The sketch is not to-scale. For simplicity, only one indirect line of sight is shown, the unit vector $\hat{n}$ describes the direction of the line of sight. The latitudinal polar coordinate, $\theta$, is used to parametrise the velocity structure in Equation \ref{paper2:eqn:velocity}.  All velocities are directed radially, $\hat{r}$.  The entry and exit points occur at the locations indicated at the intersections of line of sight with the jet wall.}
\label{fig8}
\end{figure}

\begin{figure}
\includegraphics[width=\linewidth]{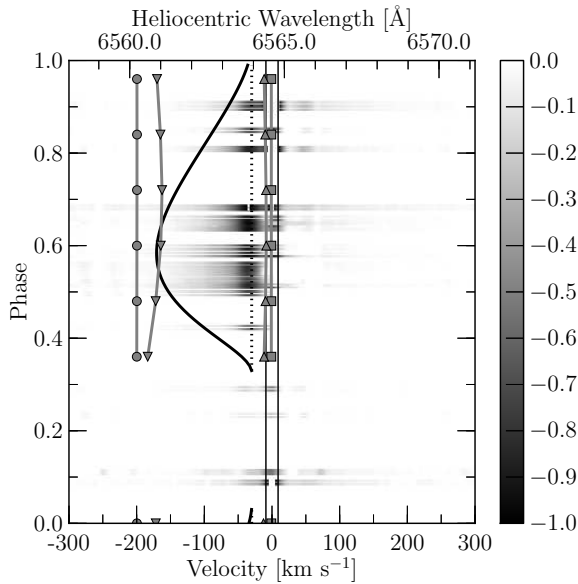}
%dynamical_color
\caption[H$\alpha$ dynamical spectrum with velocity curves]{The dynamical spectrum shown is identical to Figure \ref{fig5}. The overlain black curves correspond to best fitting model (latitudinally dependant).  The solid black curve is the projected line of sight velocity (PLOSV) along the jet axis, $v_{0}^{\mathrm{LOS}}$, while the dotted black line corresponds to the PLOSV at the entrance to the jet,$v_{\alpha}^{\mathrm{LOS}}$.  The grey curve with square markers is the $v_{0}^{\mathrm{LOS}}$ for the constant velocity model, while the grey curve with triangle markers is the $v_{0}^{\mathrm{LOS}}$ for the accelerating case.  The grey curve with circle markers is the $v_{\alpha}^{\mathrm{LOS}}$ for the constant velocity case, while the grey curve with the upside-down triangle markers is the $v_{\alpha}^{\mathrm{LOS}}$ for the accelerating model.  The range of phases from $\phi$~=~0.05 to $\phi$~=~0.3 show no curves because the indirect line of sight does not pass through the jet. }
\label{fig9}
\end{figure}

The best fitting model was the latitudinal velocity structure, which has a velocity gradient (with the higher velocities along the jet axis) in the latitudinal (polar) direction only. Typically, acceleration of the outflowing jet normally happens within 1~au for solar type stars \citep{appenzeller2005}, after which the motion is ballistic.  Since our line of sight never probes lower than roughly 1~au, see Figure \ref{fig7}, we will ignore such acceleration in this model.  Typically, the terminal speed along a given streamline scales with the escape speed at the base of the streamline where the material is launched.  For this reason hydrodynamic models predict velocity distributions with the highest velocity material along the cone axis (sometimes referred to as a spine or jet) and the slower material at the edges (sometimes called the sheath or conical wind) \citep[e.g.,][]{
romanova2009}.  The velocity of the launched material is dependent upon distance from the centre of mass of the secondary star. We model this dependency as a latitudinally dependent power law velocity structure,
\begin{equation}
 v(\theta) \hat{r}= \left[ v_{0} + \left( v_{\alpha} - v_{0}\right) \left(\frac {\theta} {\alpha}\right)^{p}\right] \hat{r} \ ,
\label{paper2:eqn:velocity}
\end{equation} where $\theta$ is the polar latitudinal coordinate (which varies from 0{\degr} to $\alpha$), $v_{0}$ is the velocity along the jet axis (set to 200~km~s$^{-1}$), $v_{\alpha}$ is the velocity at the edge of the jet (set to 30~km~s$^{-1}$). By symmetry $v(\theta)$ must be an even function of $\theta$, so we set p to an even integer. A constant velocity jet is the case $p = 0$, and for the spine sheath case we set $p = 2$. Figure \ref{fig8} illustrates a simplified system geometry indicating the unit vectors and coordinates used in this model.  As a function of orbital phase, projected line of sight velocities along the jet axis $v_{0}^{\mathrm{LOS}} = v_{0}\hat{r} \cdot \hat{n}$ and jet entrance $v_{\alpha}^{\mathrm{LOS}} = v_{\alpha}\hat{r} \cdot \hat{n}$ are over-plotted on the dynamical spectrum in Figure \ref{fig9}.  The maximum of $v_{0}^{\mathrm{LOS}}$ is roughly 160~km~s$^{-1}$ and the orbital phase dependence of the projected jet axis velocity matches the blue 
edge of the absorption feature, $V_{\mathrm{blue \ edge}}$, reasonably well. The calculations show that projected jet entrance velocity $v_{\alpha}^{\mathrm{LOS}}$ is roughly constant at 35~km~s$^{-1}$, which is where the absorption maximum $V_{\mathrm{abs \ max}}$ occurs in the data. We find that this simple model fits the data surprisingly well, matching not only the observed velocities, but the observed orbital phase dependence as well. Any arbitrary model can fit a single observation, but the fact that one model fits all of the observed data so well makes this fit reasonable.  Included in Figure \ref{fig9} are curves representing the other two models (grey curves).  The grey curves are for the same locations in the jet as the best fitting model. It is clear that the grey curves do not fit the data.

It appears that $V_{\mathrm{abs \ max}}$ and $v_{\alpha}^{\mathrm{LOS}}$ are related to each other, and that $V_{\mathrm{blue \ edge}}$ and $v_{0}^{\mathrm{LOS}}$ are also related to one another (the best fit to the data gives $v_{0} = 200$~km~s$^{-1}$ and $v_{\alpha} = 30$~km~s$^{-1}$). To understand this relation, let us look at why it might be the case. Continuum photons from the primary star can only be absorbed via a transition line (e.g., H$\alpha$) if the frequency of the photon lies within the range of frequencies spanned by the line profile.  However, in the jet the material is moving.  Therefore, the transition line in the jet must be Doppler shifted to the velocity of the absorbing particle.  The continuum photon from the star can only be absorbed if its frequency lies within the line width ($\Delta \nu_{\mathrm {D}}$)  frequencies of the Doppler shifted transition line.  The geometric length, $ds$, through the jet in which the continuum photon may be absorbed depends on the width of the line and 
the velocity gradient $(dv_{\mathrm{LOS}} / ds)$ of the moving absorbers. In the limit that the line is very narrow, the geometric path length (Sobolev length) depends only on the velocity gradient. A large gradient in the velocity of the absorbing material in the jet will lead to a narrow interaction region, while a smaller velocity gradient gives rise to a longer interaction region.  The longer the interaction region the stronger the absorption. Therefore, we expect that the absorption will be strongest near the walls of the jet,  due to the relatively small velocity gradient (d$v(\theta) \hat{r}$/d$\theta$) and correspondingly longer interaction length.  We expect that the material near the jet axis will have smaller absorption due to the larger velocity gradient, while the material near the entry point will have intermediate absorption values.  We now have a physical explanation of why $V_{\mathrm{abs \ max}}$~=~$v_{\alpha}^{\mathrm{LOS}}$, and $V_{\mathrm{blue \ edge}}$~=~$v_{0}^{\mathrm{LOS}}$.

In \citet{witt2009} it was noted that the velocity of the outflow was variable based on the observed variation of $V_{\mathrm{blue \ edge}}$, see Figure \ref{fig9}. This variability was interpreted as a jet that varies in launch speed or size as a function of orbital phase. The range in the variation of the velocity would imply that the disk is substantially changing size, since the velocity of the launched material depends on the distance from the centre of the secondary star.  Such a dramatic change in the radius of accretion disk is in contradiction to the fall-in time as calculated in \citet{witt2009} of roughly
500 years. The fact that the accretion disk is stable for longer periods of
times than the orbital period implies that the jet is likely stable
as well. Variable accretion rates could possibly affect material launched
from the edges of the disk; however, this material should be launched
with slower velocities and cannot explain the observed variation in $V_{\mathrm{blue \ edge}}$. 
In contrast, as shown by our calculations, a conical jet with a latitudinally-dependent velocity
structure and fixed half opening angle angle, $\alpha$, fits the observations remarkably well, and our model does not require a complex interpretation involving a jet that changes size or launch speed with orbital phase.  Finally, we note that neither a constant velocity jet ($p = 0$) nor a radially accelerating jet (without angular variation) can reproduce the phase dependence of the velocities of the blue edge, $V_{\mathrm{blue \ edge}}$, and the absorption maximum, $V_{\mathrm{abs \ max}}$.

\section{Discussion and conclusions}\label{paper2:sec:conclusions}

In this study we analysed the blue shifted absorption feature previously attributed to the bipolar jet emanating from a hot accretion disk around the secondary star \citep{witt2009}. This jet system shares many similarities with jet observed in BD$+$46{\degr}442 \citep{gorlova2012}.  Both systems are evolved binaries with circumbinary discs, and both systems have wide jets, which are launched from the secondary stars.  However, the viewing geometry of the two objects differ.

Based on the observed range in orbital phase during which the primary star is occulted, we find the half opening angle of the jet, $\alpha$, to be slightly less than the effective inclination angle (35{\degr}). It cannot be greater
than this angle, as the spectrum is virtually unaffected near
the phase of inferior conjunction (IC). This conclusion is independent of the assumed inclination
angle. \citet{waelkens1996} showed that they needed an effective inclination of 35{\degr} in order to
fit the observed periodic variation in the visual magnitude; however,
they used the geometry of the X-shaped bi-cone to infer this effective inclination angle. That said, since our jet half opening angle 
is essentially the same as the viewing angle it seems very likely that the observed jet is actively carving out the biconical cavity 
in the otherwise spherical outflow from the AGB stage of the primary star's life.  Furthermore, in order to fit the equivalent width 
measurements, a wide jet is required.  Therefore, our observations are inconsistent with a narrow precessing jet model, as recently 
proposed by \citet{velazquez2011}.

The velocity of the blue edge of the absorption, $V_{\mathrm{blue \ edge}}$, is seen to vary. We conclude that this variability is
 likely the result of our indirect line of sight passing through a jet with a latitudinal velocity structure, and not to a jet that
 is variable with orbital phase.  The jet is stable over the duration of the binary orbit, and does not show variation of outflow 
velocity or direction.  Over the 8 year span of observations no precession in the jet is apparent.  The maximum outflow velocity in 
the jet is $v_{0} = 200$~km~s$^{-1}$ on the jet axis, and $V_{\mathrm{blue \ edge}}$ is physically related to this velocity. The 
velocity of the absorption maximum, $V_{\mathrm{abs \ max}}$, in the jet occurs at approximately $-$35~km~s$^{-1}$, which is 
physically related to the velocity at the walls of the jet, which we find to be $v_{\alpha} = 30$~km~s$^{-1}$.

\section*{Acknowledgements}

The authors would like to acknowledge the efforts of APO observers who helped acquire the data.  We would also like to thank Hans van Winckel for his constructive referee report.

\bsp \label{lastpage}


\begin{thebibliography}{99}

% A %%%%%%%%%%%%%%%%%%%%%%%%%%%%%%%%%%%%%%%%%%%%%%%%%%

\bibitem[\protect\citeauthoryear{Allard 
\& Hauschildt}{1995}]{allardhau1995} Allard F., Hauschildt P.~H., 1995, ApJ, 445, 433 

\bibitem[\protect\citeauthoryear{Allard et al.}{2001}]{allard2001} 
Allard F., Hauschildt P.~H., Alexander D.~R., Tamanai A., Schweitzer A., 
2001, ApJ, 556, 357 

\bibitem[\protect\citeauthoryear{Appenzeller et al.}{2005}]{appenzeller2005} Appenzeller I., Bertout C., Stahl O., 2005, A\&A, 434, 
1005 

% B %%%%%%%%%%%%%%%%%%%%%%%%%%%%%%%%%%%%%%%%%%%%%%%%%%

\bibitem[\protect\citeauthoryear{Baron et al.}{1996}]{baron1996} 
Baron E., Hauschildt P.~H., Nugent P., Branch D., 1996, MNRAS, 283, 297 

\bibitem[\protect\citeauthoryear{Baron 
\& Hauschildt}{1998}]{baronhau1998} Baron E., Hauschildt P.~H., 1998, ApJ, 495, 370 

\bibitem[\protect\citeauthoryear{Bujarrabal et 
al.}{2005}]{bujarrabal2005} Bujarrabal V., Castro-Carrizo A., Alcolea J., Neri R., 2005, A\&A, 441, 1031 

% C %%%%%%%%%%%%%%%%%%%%%%%%%%%%%%%%%%%%%%%%%%%%%%%%%%

\bibitem[\protect\citeauthoryear{Cohen et al.}{1975}]{cohen1975} 
Cohen M., et al., 1975, ApJ, 196, 179 

\bibitem[\protect\citeauthoryear{Cohen et al.}{2004}]{cohen2004}
Cohen M., van Winckel H., Bond H.~E., Gull T.~R., 2004, AJ, 127, 2362 


% D %%%%%%%%%%%%%%%%%%%%%%%%%%%%%%%%%%%%%%%%%%%%%%%%%%
% G %%%%%%%%%%%%%%%%%%%%%%%%%%%%%%%%%%%%%%%%%%%%%%%%%%

\bibitem[\protect\citeauthoryear{Gorlova et 
al.}{2012}]{gorlova2012} Gorlova N., et al., 2012, A\&A, 542, A27 


\bibitem[\protect\citeauthoryear{Grinin, Mitskevich, 
\& Tambovtseva}{2006}]{grinin2006} Grinin V.~P., Mitskevich A.~S., Tambovtseva L.~V., 2006, AstL, 32, 110G

\bibitem[\protect\citeauthoryear{Grinin, Tambovtseva, 
\& Weigelt}{2012}]{grinin2012} Grinin V.~P., Tambovtseva L.~V., Weigelt G., 2012, A\&A, 544, A45

% H %%%%%%%%%%%%%%%%%%%%%%%%%%%%%%%%%%%%%%%%%%%%%%%%%%

\bibitem[\protect\citeauthoryear{Hauschildt}{1992}]{hau1992} 
Hauschildt P.~H., 1992, JQSRT, 47, 433 

\bibitem[\protect\citeauthoryear{Hauschildt}{1993}]{hau1993} 
Hauschildt P.~H., 1993, JQSRT, 50, 301 

\bibitem[\protect\citeauthoryear{Hauschildt 
\& Baron}{1995}]{hau1995} Hauschildt P.~H., Baron E., 1995, JQSRT, 54, 987 

\bibitem[\protect\citeauthoryear{Hauschildt, Baron, 
\& Allard}{1997}]{hau1997} Hauschildt P.~H., Baron E., Allard F., 1997, ApJ, 483, 390 

\bibitem[\protect\citeauthoryear{Hauschildt et 
al.}{1996}]{hau1996} Hauschildt P.~H., Baron E., Starrfield S., 
Allard F., 1996, ApJ, 462, 386 

\bibitem[\protect\citeauthoryear{Hauschildt, Lowenthal, 
\& Baron}{2001}]{hau2001} Hauschildt P.~H., Lowenthal D.~K., Baron E., 2001, ApJS, 134, 323 


\bibitem[\protect\citeauthoryear{Hobbs et al.}{2004}]{hobbs2004}
Hobbs L.~M., Thorburn J.~A., Oka T., Barentine J., Snow T.~P., York D.~G., 
2004, ApJ, 615, 947 

% I %%%%%%%%%%%%%%%%%%%%%%%%%%%%%%%%%%%%%%%%%%%%%%%%%%

% J %%%%%%%%%%%%%%%%%%%%%%%%%%%%%%%%%%%%%%%%%%%%%%%%%%
\bibitem[\protect\citeauthoryear{Jura et al.}{1997}]{jura1997} Jura M., Turner J., Balm S.~P., 1997, ApJ, 474, 741 

% K %%%%%%%%%%%%%%%%%%%%%%%%%%%%%%%%%%%%%%%%%%%%%%%%%%

\bibitem[\protect\citeauthoryear{Koning et al.}{2011}]{koning2011} Koning N., Kwok S., Steffen W., 2011, ApJ, 740, 27 

% \bibitem[\protect\citeauthoryear{Kovalev et 
% al.}{2007}]{kovalev2007} Kovalev Y.~Y., Lister M.~L., Homan D.~C., 
% Kellermann K.~I., 2007, ApJ, 668, L27 

% L %%%%%%%%%%%%%%%%%%%%%%%%%%%%%%%%%%%%%%%%%%%%%%%%%%

% M %%%%%%%%%%%%%%%%%%%%%%%%%%%%%%%%%%%%%%%%%%%%%%%%%%

\bibitem[\protect\citeauthoryear{Men'shchikov et 
al.}{2002}]{menshchikov2002} Men'shchikov A.~B., Schertl D., Tuthill P.~G., Weigelt G., Yungelson L.~R., 2002, A\&A, 393, 867 

\bibitem[\protect\citeauthoryear{Morris}{1981}]{morris1981} Morris 
M., 1981, ApJ, 249, 572 

\bibitem[\protect\citeauthoryear{Morris}{1987}]{morris1987} Morris 
M., 1987, PASP, 99, 1115 

\bibitem[\protect\citeauthoryear{Morton}{1974}]{morton1974} Morton 
D.~C., 1974, ApJ, 193, L35 

% O %%%%%%%%%%%%%%%%%%%%%%%%%%%%%%%%%%%%%%%%%%%%%%%%%%


\bibitem[\protect\citeauthoryear{Osterbart et al.}{1997}]{osterbart1997} Osterbart R., Langer N., Weigelt G., 1997, A\&A, 325, 609 

% P %%%%%%%%%%%%%%%%%%%%%%%%%%%%%%%%%%%%%%%%%%%%%%%%%%

% R %%%%%%%%%%%%%%%%%%%%%%%%%%%%%%%%%%%%%%%%%%%%%%%%%%
% 

\bibitem[\protect\citeauthoryear{Roddier et 
al.}{1995}]{roddier1995} Roddier F., Roddier C., Graves J.~E., 
Northcott M.~J., 1995, ApJ, 443, 249 

\bibitem[\protect\citeauthoryear{Romanova et 
al.}{2009}]{romanova2009} Romanova M.~M., Ustyugova G.~V., Koldoba 
A.~V., Lovelace R.~V.~E., 2009, MNRAS, 399, 1802 


% S %%%%%%%%%%%%%%%%%%%%%%%%%%%%%%%%%%%%%%%%%%%%%%%%%%


\bibitem[\protect\citeauthoryear{Schmidt et al.}{1980}]{schmidt1980} Schmidt G.~D., Cohen M., Margon B., 1980, ApJ, 239, L133 

\bibitem[\protect\citeauthoryear{Schmidt 
\& Witt}{1991}]{schmidt1991} Schmidt G.~D., Witt A.~N., 1991, ApJ, 383, 698 


\bibitem[\protect\citeauthoryear{Soker}{2005}]{soker2005} Soker N., 2005, AJ, 129, 947 

% T %%%%%%%%%%%%%%%%%%%%%%%%%%%%%%%%%%%%%%%%%%%%%%%%%%

\bibitem[\protect\citeauthoryear{Thomas et al.}{2011}]{thomas2011} 
Thomas J.~D., et al., 2011, MNRAS, 417, 2860 

% \bibitem[\protect\citeauthoryear{Thomas}{2012}]{thomas2012} 
% Thomas J.~D., 2012, Dissertation --University of Toledo, Toledo OH 


\bibitem[\protect\citeauthoryear{Thorburn et 
al.}{2003}]{thorburn2003} Thorburn J.~A., et al., 2003, ApJ, 584, 
339

% V %%%%%%%%%%%%%%%%%%%%%%%%%%%%%%%%%%%%%%%%%%%%%%%%%%

\bibitem[\protect\citeauthoryear{van Winckel et al.}{1995}]{vanwinckel1995} van Winckel H., Waelkens C., Waters L.~B.~F.~M., 1995, A\&A, 293, L25 

\bibitem[\protect\citeauthoryear{Vel{\'a}zquez et 
al.}{2011}]{velazquez2011} Vel{\'a}zquez P.~F., Steffen W., Raga 
A.~C., Haro-Corzo S., Esquivel A., Cant{\'o} J., Riera A., 2011, ApJ, 734, 
57 

% \bibitem[\protect\citeauthoryear{Venn 
% \& Lambert}{1990}]{venn1990} Venn K.~A., Lambert D.~L., 1990, ApJ, 363, 234 
% 
% 
% \bibitem[\protect\citeauthoryear{Vijh, Witt, 
% \& Gordon}{2004}]{vijh2004} Vijh U.~P., Witt A.~N., Gordon K.~D., 2004, ApJ, 606, L65 
% 
% \bibitem[\protect\citeauthoryear{Vijh, Witt, 
% \& Gordon}{2005}]{vijh2005} Vijh U.~P., Witt A.~N., Gordon K.~D., 2005, ApJ, 633, 262 

\bibitem[\protect\citeauthoryear{Vijh et al.}{2006}]{vijh2006} 
Vijh U.~P., Witt A.~N., York D.~G., Dwarkadas V.~V., Woodgate B.~E., 
Palunas P., 2006, ApJ, 653, 1336 


% W %%%%%%%%%%%%%%%%%%%%%%%%%%%%%%%%%%%%%%%%%%%%%%%%%%


\bibitem[\protect\citeauthoryear{Waelkens et 
al.}{1996}]{waelkens1996} Waelkens C., van Winckel H., Waters L.~B.~F.~M., Bakker E.~J., 1996, A\&A, 314, L17 

\bibitem[\protect\citeauthoryear{Wang et al.}{2003}]{wang2003} 
Wang X., Wang B., Pouch J., Miranda F., Fisch M., Anderson J.~E., Sergan V., Bos P.~J., 2003, SPIE, 5162, 139 

\bibitem[\protect\citeauthoryear{Witt 
\& Boroson}{1990}]{witt1990} Witt A.~N., Boroson T.~A., 1990, ApJ, 355, 182 

\bibitem[\protect\citeauthoryear{Witt et al.}{2006}]{witt2006} 
Witt A.~N., Gordon K.~D., Vijh U.~P., Sell P.~H., Smith T.~L., Xie R.-H., 
2006, ApJ, 636, 303 

\bibitem[\protect\citeauthoryear{Witt et al.}{2009}]{witt2009} 
Witt A.~N., Vijh U.~P., Hobbs L.~M., Aufdenberg J.~P., Thorburn J.~A., York 
D.~G., 2009, ApJ, 693, 1946 

%%%%%%%%%%%%%%%%%%%%%%%%%%%
%%%%%%%%%%%%%%%%%%%%%%%%%%%


\end{thebibliography}
\end{document}